\documentclass[prl,twocolumn,floatfix,firstinits,superscriptaddress]{revtex4-1}

\usepackage{amsmath}
\usepackage{amssymb}
\usepackage{float}
\usepackage{subfig}
\usepackage{multirow}

\setcitestyle{square}
\usepackage[font=small,skip=4pt, justification=raggedright,format=plain,singlelinecheck=false]{caption}

\usepackage{graphicx}
\graphicspath{{../FIGURE/}}
\usepackage{sidecap}

\usepackage[usenames,dvipsnames]{color}
\usepackage[colorlinks=true,urlcolor=Blue,citecolor=Violet,linkcolor=BrickRed]{hyperref}

\newcommand{\Dapp}{D_{\rm app}}

\begin{document}

\title{What Does FEXI Measure?}
%------------------------------
\author{Mohammad Khateri}
\affiliation{A.I.Virtanen Institute for Molecular Sciences, University of Eastern Finland, Kuopio, Finland}
\author{Marco Reisert}
\affiliation{Medical Physics, Department of Radiology, University Medical Center Freiburg, Faculty of Medicine, University of Freiburg, Freiburg, Germany}
\affiliation{Department of Stereotactic and Functional Neurosurgery, University Medical Center Freiburg, Faculty of Medicine, University of Freiburg, Freiburg, Germany}
\author{Alejandra Sierra}
\author{Jussi Tohka}
\affiliation{A.I.Virtanen Institute for Molecular Sciences, University of Eastern Finland, Kuopio, Finland}
\author{Valerij G.\ Kiselev}
\email[Corresponding author: ]{kiselev@ukl.uni-freiburg.de}
\affiliation{Medical Physics, Department of Radiology, University Medical Center Freiburg, Faculty of Medicine, University of Freiburg, Freiburg, Germany}

\begin{abstract}\noindent
Filter-exchange imaging (FEXI) has already been utilized in several biomedical studies for evaluating the permeability of cell membranes. The method relies on suppressing the extracellular signal using strong diffusion weighting (the mobility filter causing a reduction in the overall diffusivity) and monitoring the subsequent diffusivity recovery. Using Monte Carlo (MC) simulations, we  demonstrate that FEXI is not uniquely sensitive to the transcytolemmal exchange but also to the geometry of involved compartments: Complex geometry offers locations where spins remain unaffected by the mobility filter; moving to other locations afterward, such spins contribute to the diffusivity recovery without actually permeating any membrane. This exchange mechanism warns those who aim to use FEXI in complex media such as brain gray matter and opens large room for investigation towards crystallizing the genuine membrane permeation and characterizing the compartment geometry.
\end{abstract}

\maketitle

%------------------------------

\raisebox{98mm}[0pt][0pt]{\color[rgb]{0.5,0.5,0.8}This is a preprint of a paper already published in \href{https://analyticalsciencejournals.onlinelibrary.wiley.com/doi/10.1002/nbm.4804}{NMR in Biomedicine}. When \href{https://analyticalsciencejournals.onlinelibrary.wiley.com/action/showCitFormats?doi=10.1002\%2Fnbm.4804}{cite} it, please, refer to the \href{https://analyticalsciencejournals.onlinelibrary.wiley.com/doi/10.1002/nbm.4804}{journal version}.}

%------------------------------

\section{Introduction}\label{sec1}

\begin{figure*}[tbp]
\includegraphics[width=0.96\textwidth]{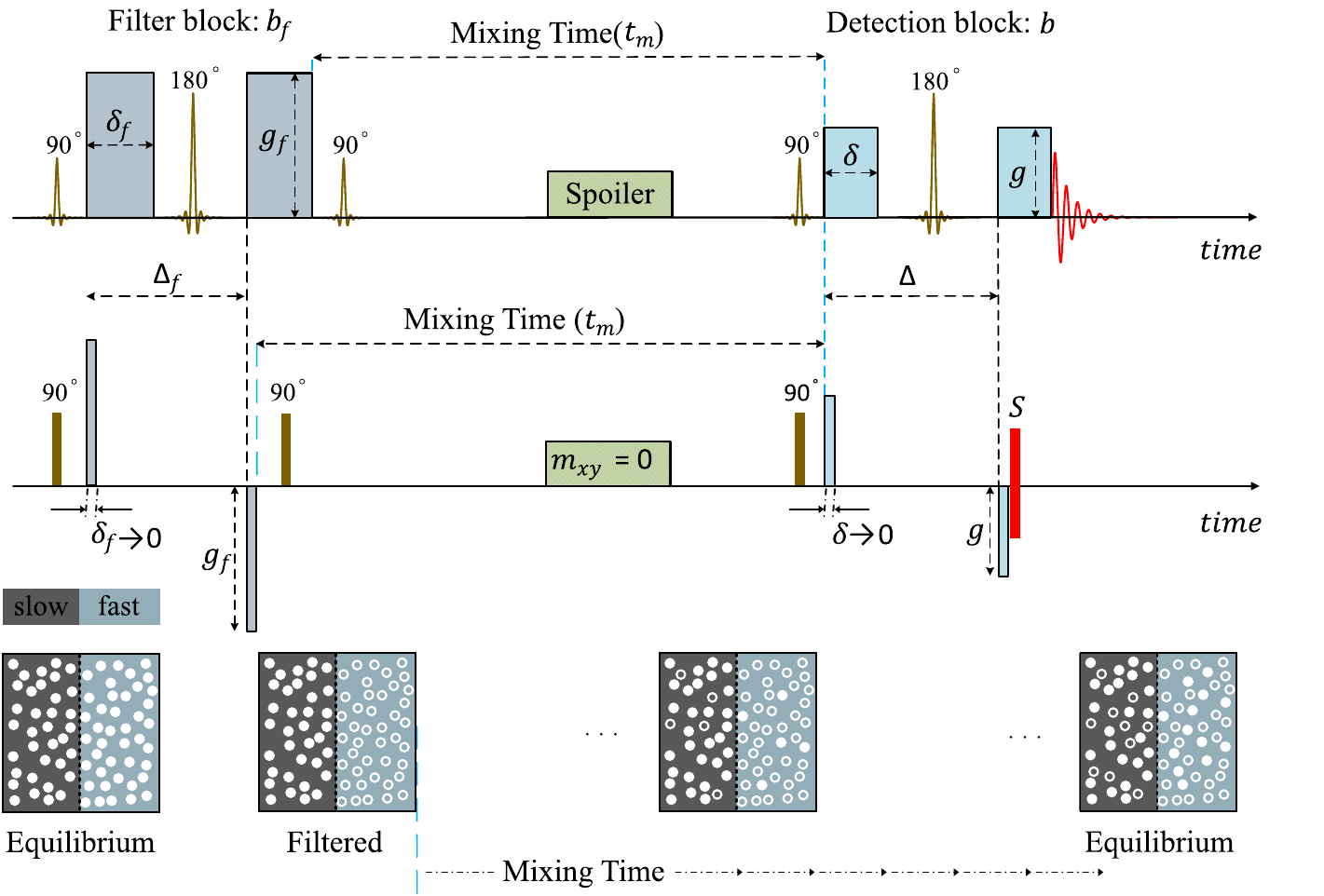}
\caption{\footnotesize Schematics of FEXI (upper row) and how it is simulated in this study (middle row). FEXI concatenates a filter block and a detection block separated by a mixing time $t_m$. These two blocks are specified by diffusion sensitizing gradients {$g_f$ and $g$}, duration ({$\delta_f$} and {$\delta$}), and diffusion time ({$\Delta_f$} and {$\Delta$}), respectively. The derived strength of diffusion weighting is $b_f=(\gamma g_f \delta_f)^2(\Delta_f-\delta_f/3)$ and {$b=(\gamma g \delta)^2(\Delta-\delta/3)$}, respectively. The bottom row illustrates how FEXI works on  a two-site system, including compartments with fast and slow diffusion marked by the light gray and dark gray background, respectively. Solid circles stand for molecules contributing to the signal, empty circles for the signal suppressed by the filter. From left to right: (i) Water molecules are in the equilibrium; (ii) mobility filter suppresses signal in the compartments with fast diffusion; (iii) during mixing time, water molecules exchange and gradually restore the equilibrium between compartments; and, (iv) detection block measures the apparent diffusion coefficient $\Dapp(t)$ through the mixing time. The recovery of $\Dapp(t)$ towards the equilibrium can be interpreted in terms of molecule permeation through cell membranes \cite{qiao2005diffusion} and geometry of compartments \cite{bernin2013nmr}. 
}
\label{FEXI_sequence}
\end{figure*}

Diffusion-weighted NMR offers several ways for evaluating exchange in samples with nontrivial microstructure. As a classical correlation technique, diffusion exchange spectroscopy (DEXSY) \cite{callaghan2004diffusion,schillmaier2020disentangling} reveals populations of molecules that experience a change in their diffusion environment during a certain mixing time. A more parsimonious double pulsed-gradient spin echo (PGSE) measurement focuses on the signal dependence on the angle between the directions of two gradient pairs \cite{Mitra95,Shemesh2016}. Modification of this technique to have gradients pairs of different magnitude  gave rise to the filter-exchange imaging (FEXI) \cite{aaslund2009filter} for mapping the apparent exchange rate (AXR) between compartments with different diffusivities (Figure \ref{FEXI_sequence}).  The first gradient pair acts as a mobility filter suppressing the signal from the compartment with high diffusivity. The second gradient pair measures the diffusion coefficient for different mixing times, revealing the recovery to the unperturbed value. The specific exchange mechanism is not crucial for the functioning of the method. Its precursor, the diffusion -- diffusion correlation measurement was implemented for liquid crystals \cite{callaghan2004diffusion} and microporous materials \cite{gratz2009multidimensional} where the exchange was mediated by diffusion between different microenvironments.  The present revival of interest in such measurements is inspired by possible biomedical applications. The method was validated in numerical simulations \cite{lasivc2011apparent,ludwig2021apparent}, yeast cell suspension \cite{aaslund2009filter,lasivc2011apparent} and human embryonic kidney cells \cite{schilling2017mri}. Implemented on clinical magnetic resonance imaging (MRI) scanners \cite{lasivc2011apparent}, it has been applied to investigation of human brain tissue \cite{nilsson2013noninvasive,soenderby2014apparent,bai2020feasibility,li2022direction}, brain tumors \cite{lampinen2017optimal} and breast cancer \cite{lasivc2016apparent}, {see also an available review \cite{bernin2013nmr}}. While the multiplicity of exchange mechanisms is well known \cite{bernin2013nmr,nilsson2013noninvasive}, the interpretation of measurements often focuses on solely the transcytolemmal water exchange.

%%%%%%%%%%%%%%%%%%%

%Fig1---------------------

%Fig2---------------------
\begin{figure*}[tbp]
\includegraphics[width = 0.90\textwidth]{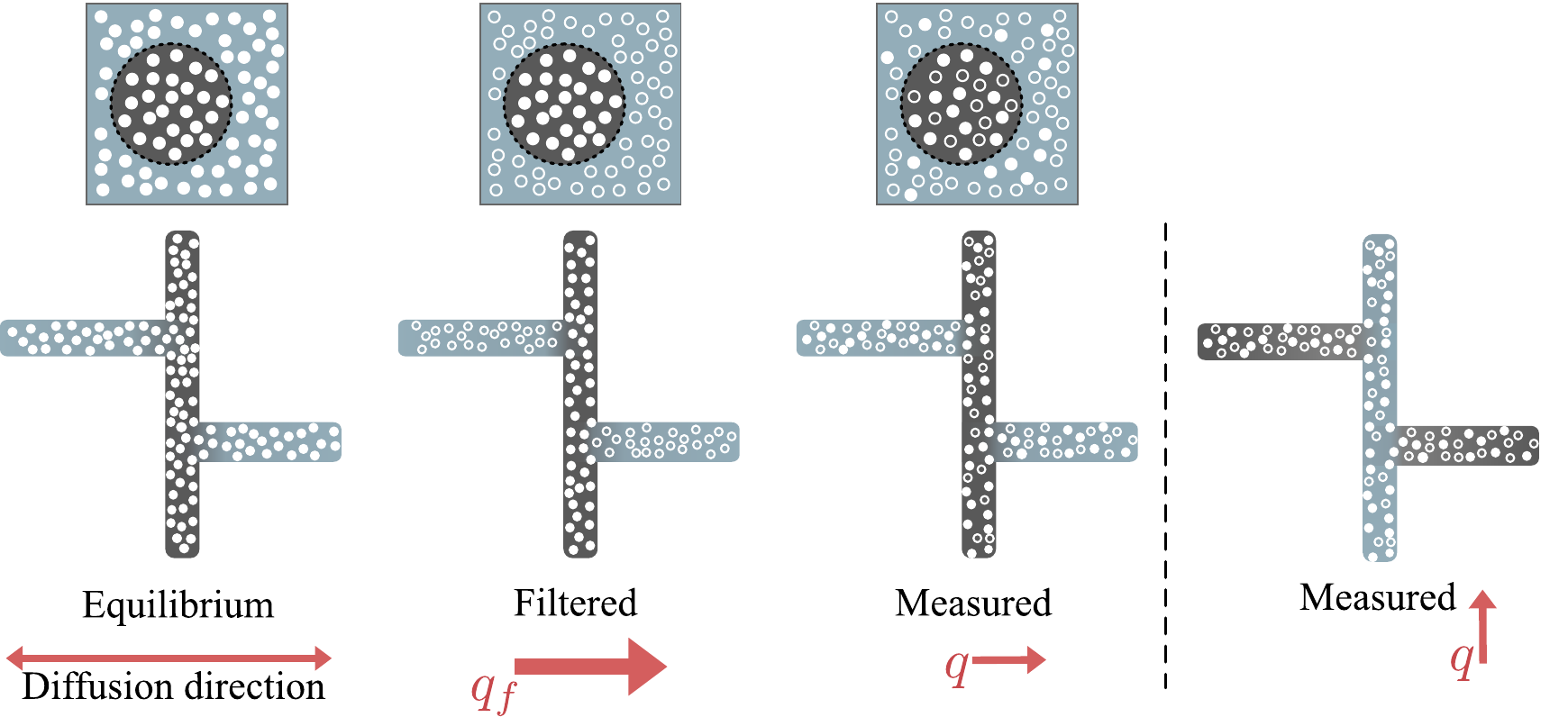}
\caption{\footnotesize Illustration of inter-and intra-compartment exchange mechanisms. (I) transcytolemmal, an inter-compartment exchange, occurs due to the exchange of the molecules through the permeable boundaries of closed compartments (first row) \cite{qiao2005diffusion}. And (II) diffusion-mediated exchange (second row), an intra-compartment exchange between anisotropic domains with different orientations\cite{bernin2013nmr}; some water escapes suppression by the mobility filter in locations with small size in the direction of the filter gradient. The following diffusional motion results in exchange that actually occurs within the same compartment with no permeation through any membrane. Note that the fast and slow are not absolute being defined by the gradient direction.}
\label{fig_exchange}
\end{figure*}

%Fig3---------------------
\begin{figure*}[tbp]
\includegraphics[width=0.92\textwidth]{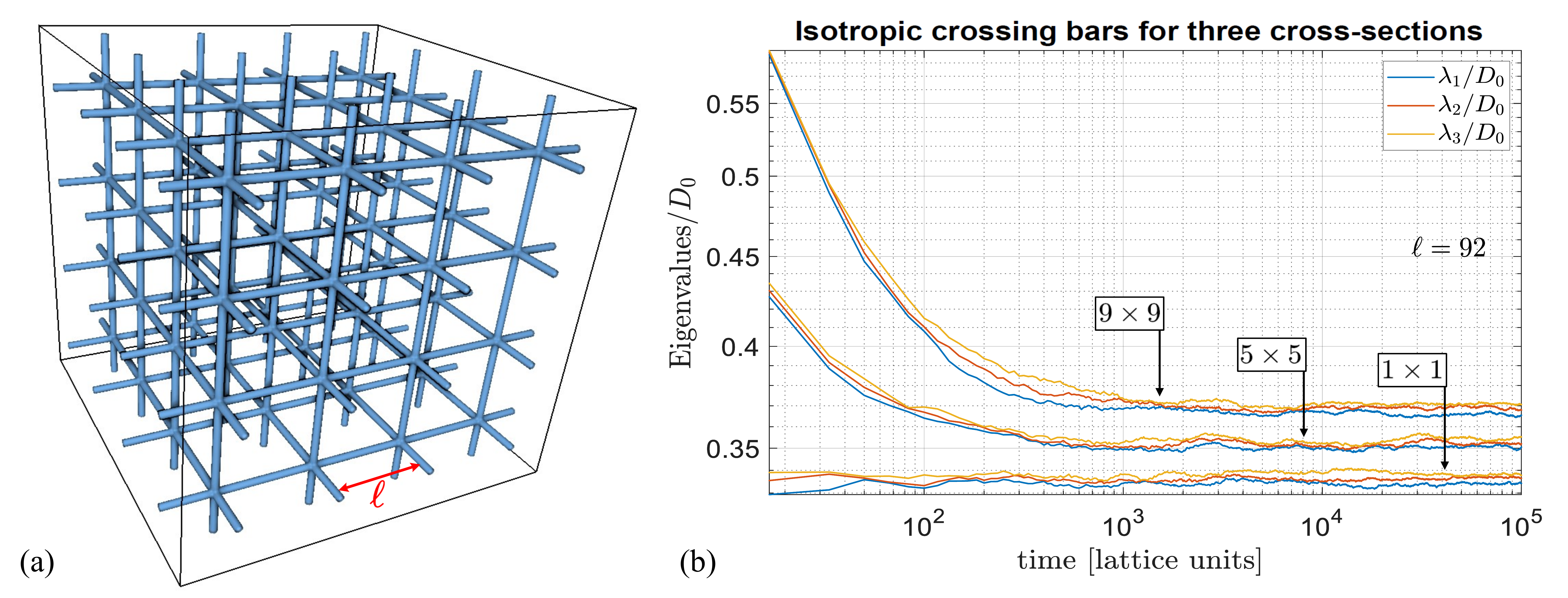}
\caption{\footnotesize (a) A synthetic medium with a regular geometry used to illustrate the diffusion-mediated exchange mechanism. It is constructed with three regular arrays of parallel bars oriented in three orthogonal directions. The spacing between the bars in each group is referred to as the segment length $\ell$. (b) The time-dependent eigenvalues of diffusion tensors, Equation \ref{def_Dab} for ``intra-cellular'' diffusion inside the bars with the cross-sections $1\times1$, $5\times5$, and $9\times9$ voxel sides, and the fixed $\ell=92$ voxel sides. $D_0$ is the bulk diffusion coefficient.}

\label{fig_media_bar}
\end{figure*}

%Fig4---------------------
\begin{figure*}[tbp]
\centering{
\includegraphics[width=0.67\textwidth]{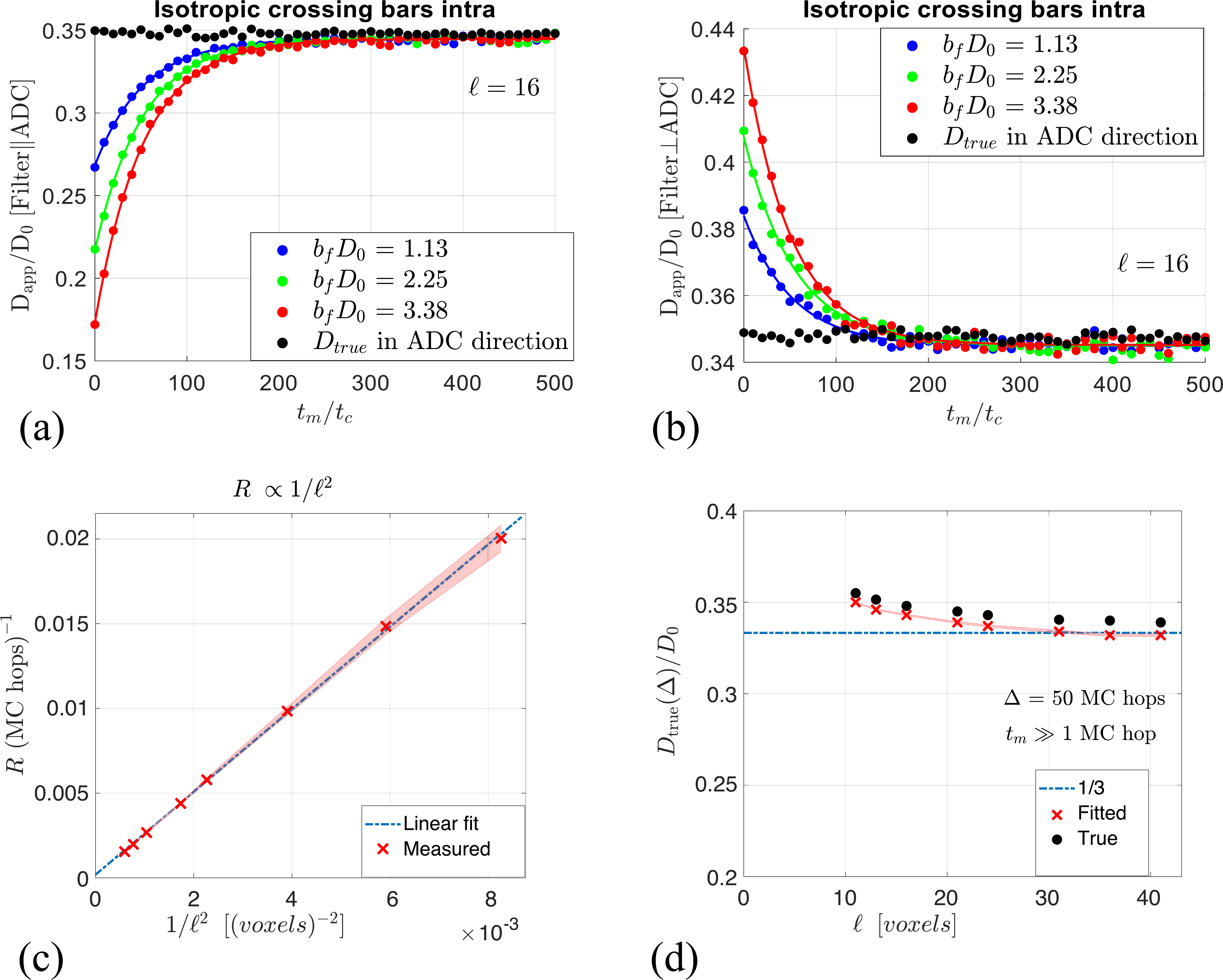}
\caption{\footnotesize  Simulated FEXI imprint of diffusion-mediated exchange for diffusion inside the bars in the medium shown in Figure \ref{fig_media_bar} with the bar cross-section $1\times 1$ voxel side. Solid lines in the panel (a,b) show Equation \ref{def_fit} fitted to data, and the transparent red colors in panel (c,d) indicate $95\%$ confidence bounds for the fitted parameters (confidence bound in panel (d) is thinner than the symbol size). $D_{\rm true}(\Delta)$ is true diffusion coefficient, Equation \ref{def_Dab} in the same direction as $\Dapp(t_m)$ and for the same diffusion time $\Delta$ as used in the FEXI measurement block (Figure \ref{FEXI_sequence}). Panel (a) shows results for the same directions of the mobility filter and the measurement gradients, panel (b) presents results when these directions are perpendicular. Panel (c) shows the fitted apparent exchange rate $R$, Equation \ref{def_fit}. The proportionality to {$1/{\ell}^2$} reflects the anticipated scaling as discussed in the text. Panel (d) shows the genuine and fitted $D_{\rm true}(\Delta)$ from Equation \ref{def_fit} as the function of segment length $\ell$ in the isotropic crossing bars when mobility filter and measurement are in the same directions. As expected, $D_{\rm true}(\Delta)$ converges to $D_0/3$ for large segment length.}
\label{fig_media_bar_results}
}
\end{figure*}

%Fig5---------------------
\begin{figure*}[tbp]
\includegraphics[width=0.98\textwidth]{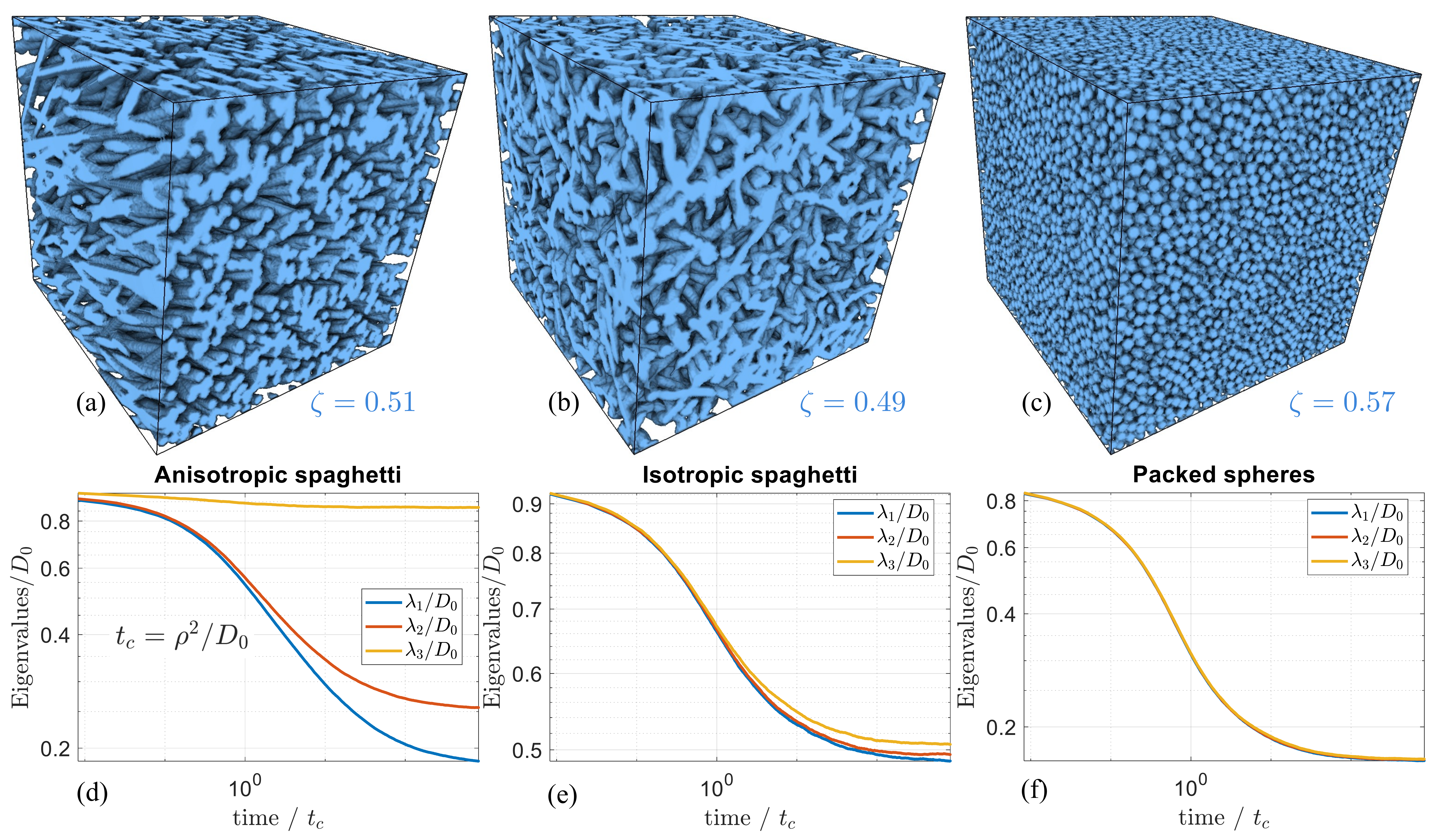}
\caption{\footnotesize  Synthetic media used in this study, each obeying periodic boundary conditions in three spatial directions with the values of $\zeta$, the fraction of  ``intracellular'' volume  (blue). (a) Anisotropic ``spaghetti'' medium formed by a randomized, self-crossing tube. (b) Isotropic ``spaghetti'' medium. (c) A medium formed by random packing of identical spheres. (d-f) The time-dependent eigenvalues of diffusion tensors for ``intra-cellular'' diffusion in corresponding media. The leveling-off in panel (f) indicates percolation between touching spheres.}
\label{fig_media}
\end{figure*}

%Fig6---------------------
\begin{figure*}[tbp]
\includegraphics[width = 0.98\textwidth]{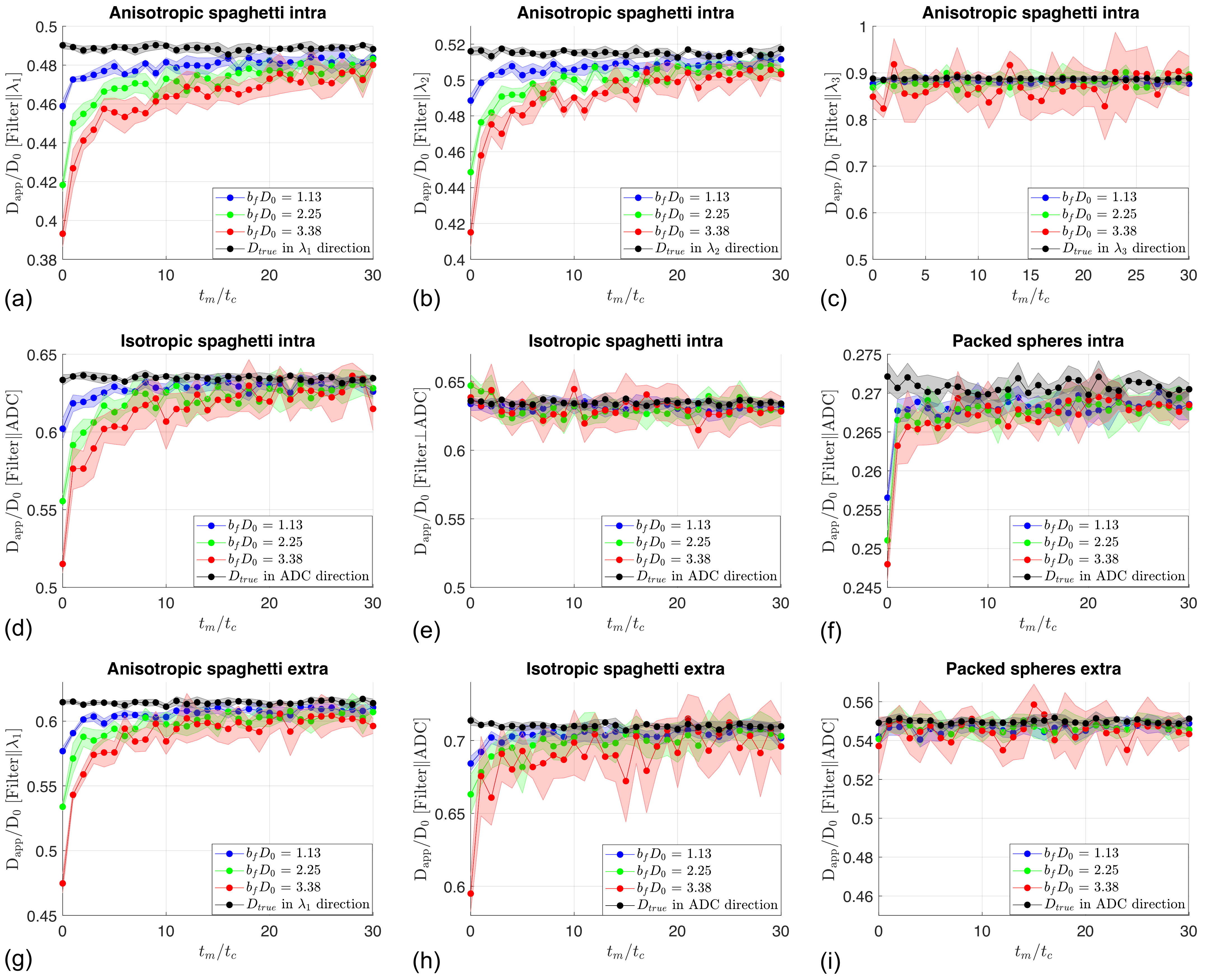}
\caption{\footnotesize  Simulation results for diffusion in media shown in Figure \ref{fig_media}, color bands at one standard deviation, {$\Delta=1.5t_c$}, where $t_c$ is defined in Equation \ref{def_tc}. Panels (a--f) show results for ``intra-cellular'' space (inside tubes or spheres), panels (g--i) for ``extra-cellular'' space. The directions of mobility filter and the measurement are parallel for all panels, but (e) for which they are orthogonal.  The label ADC and $\lambda_n$ refer to the direction of the measurement gradients coinciding with the $x$-direction of the simulation box for isotropic media and to the corresponding eigenvector for anisotropic media. A pronounced effect of diffusion-mediated exchange is seen for measurements in the most complex geometries, which are anisotropic spaghetti across the principal diffusion direction (a,b,g) and the isotropic one (d,h). Little or no effect is observed for the orthogonal direction of the filter and measurement gradients (e), along the principal diffusion direction (c) and for packed spheres (f,i).}
\label{fig_results}
\end{figure*}

%Fig7-------------------
\begin{figure}[t]
\includegraphics[width = 0.4\textwidth]{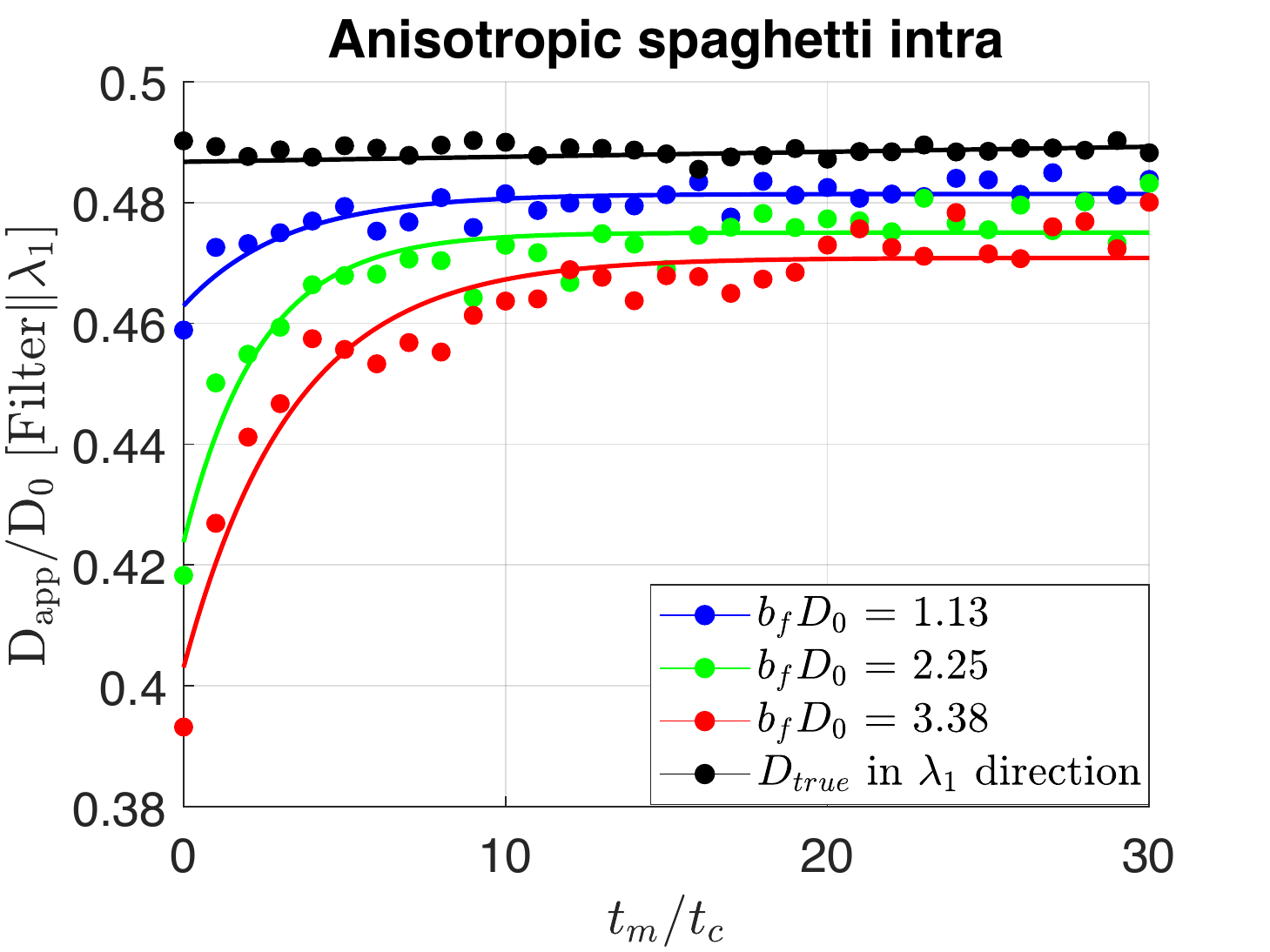}
\caption{\footnotesize The same data as in Figure \ref{fig_results}a with added lines showing Equation \ref{def_fit}\cite{lasivc2011apparent} with fitted parameters $a$, $R$ and $D_{\rm true}(\Delta)$. Affected by the slow approach of data to the asymptote, the fitting results in different $D_{\rm true}(\Delta)$ for different $b_f$. Also noticeable is the systematic deviation of fitted lines from the data.}
\label{fig_results_fit}
\end{figure}
%%%%%%%%%%%%%%%%%%

In this work, we draw attention to the need to account for diffusion-mediated exchange when interpreting FEXI in microcompartments with complex geometry. In such compartments, water can avoid suppression by the mobility filter when located in narrow spaces with limited mobility in the corresponding direction. During the mixing time, it can move to  less restricted locations where it contributes to a high diffusion coefficient (Figure \ref{fig_exchange}) . This can occur without crossing any membrane. To illustrate this mechanism, we first use Monte Carlo (MC) simulations in well-organized geometry (Figure \ref{fig_media_bar}) with qualitatively predictable results shown in Figure \ref{fig_media_bar_results}. Then, we use MC simulations in large compartments of simple to complex geometries with impermeable boundaries, shown in Figure \ref{fig_media}. While these substrates do not include any permeable membranes, our simulation results show the FEXI-typical recovery pattern of diffusion coefficient, which can be easily misinterpreted in terms of trancytolemmal exchange (Figure \ref{fig_results}). 

\section{Methods}

\subsection{Synthetic Media}
We generated a set of three-dimensional media with variable geometrical complexity. All media are described by the indicator function {$v(\bf r)$} taking the values $v=1$ within the created compartment, which we refer to as ``cellular'', and $v=0$ otherwise. 

Media with simple geometry were implemented using three regular arrays of parallel bars with the separation $\ell$ between the centers, each array aligned with one of three orthogonal directions (Figure \ref{fig_media_bar}). Three-directional crossing points formed a cubic grid with the spacing {$\ell$}, which we refer to as the segment length. The simulation boxes were mapped to the {$256^3$} lattice for {$\ell=16$}. For other $\ell$ ranging from 11 to 41, the lattice size was adjusted to satisfy the periodic boundary conditions at the simulation box faces.

Media with more complex, random geometry were generated using the algorithm previously developed for simulating the capillary network  \cite{Novikov2018_ackerman}. Each medium consists of a single object, ``the tube'' with {$v(\bf{r})=1$} obtained as the trajectory of a moving small sphere with ballistic and random velocity components. It has one beginning and one end and is otherwise continuous with periodic boundary conditions in all three directions. When generated, the trajectory had no memory about already visited locations, which lead to multiple self-crossings. We refer to this object as the ``spaghetti''. The trajectories initially generated in continuous space were mapped on a cubic lattice with the size {$256^3$}. The tube diameter, {$2\rho$} was 9 voxel sides. We used two instances of such media, one strongly anisotropic with the volume fraction {$\zeta=0.51$} (calculated as the lattice mean of {$v(\bf{r})$}) and one isotropic with {$\zeta=0.49$}, see Figure \ref{fig_media}a--b. These media differ from the known model of diffusion on piece-wise straight segments with and without branching \cite{callaghan1993principles} in the finite tube diameter, continuous curvature, and self-crossing in place of branching.

We also generated media consisting of identical, densely packed spheres with the value {$v(\bf{r})=1$} inside, positioned using a collision-driven algorithm \cite{Skoge2006} downloaded from the authors' website \cite{collisiondriven}. These media are also periodic in all three directions. Initially generated in continuous space, the media were mapped on a cubic lattice with the size {$256^3$}. Figure \ref{fig_media}c shows a medium consisting of spheres with the diameter $2\rho$ equal $9$ voxel sides and volume fraction $\zeta=0.57$.

\subsection{Monte Carlo Simulations}

For simulation, we used a C++ program, which was previously developed for simulating transverse relaxation, diffusion, and the Larmor frequency shift \cite{Novikov2018_ackerman,Novikov2008,Novikov2010,Ruh2018_size,Ruh2019_aniso}. The algorithm is based on the nearest-neighbor hopping on a cubic lattice. At each time moment, a random walker (the simulated spin) moves, with equal probability, in one of six directions. This motion is subjected to the periodic conditions at the boundaries of the simulation box. Impermeable interfaces are detected as the change in the indicator function {$v(\bf{r})$}. Steps across such interfaces are rejected and the next attempt to move is made after the time increment. 

In this work, we simulated the diffusion-weighting gradients in the narrow-pulse limit as a phase instantly acquired according to the spin's current displacement from its initial position. This displacement is the total path on the torus formed by the periodic boundary conditions (for example, making a ``rounding the world'' in one of the main directions results in the path equal the size of the simulation box). The zero point of the gradient-induced Larmor frequency shift is set to the initial spin's position. This means that each spin has its own zero-frequency point. It does not lead to any problem because the spin's signal contribution does not depend on this point for the balanced gradients. The gradients and the instant rotations simulating the radiofrequency pulses are applied according to the scheme shown in Figure \ref{FEXI_sequence}. To focus on the diffusion effect, we do not assign any transverse relaxation or magnetic susceptibility associated with {$v({\bf r})=1$}. The simulation outcome was the genuine cumulative diffusion tensor and the simulated MR signal for multiple combinations of the FEXI parameters. The diffusion tensor was calculated via the spin's displacements, ${\bf r}-{\bf r}^{(0)}$, 
\begin{eqnarray}\label{def_Dab}
D_{ab} = \frac{1}{2t} \langle [r_a-r_a^{(0)}][r_b-r_b^{(0)}]  \rangle \,
\end{eqnarray}
where {$a,b=1,2,3$} label the spatial directions, {$t$} is the time of diffusion and the averaging is performed over all spins. All simulations were performed on a computer with 2 $\times$ Intel Xeon E5 2630 CPU @2.4 GHz, and 512 GB RAM. The number of spins was set to {$10^5$}. 

The simulated signal was processed in MatLab for calculating the diffusion coefficients and result presentation. Results for random media are presented in relative units in which distances are measured in the tube or sphere radius {$\rho$} and time in the units of correlation time,
\begin{eqnarray}\label{def_tc}
t_c=\frac{\rho^2}{D_0}\,
\end{eqnarray}
where $D_0$ is the bulk diffusion coefficient. In these units, diffusion time for the mobility filter was {$\Delta_f\in\{0.5, 1, 1.5\}t_c $} and for detection block fixed to {$\Delta=1.5 t_c$}. The mixing time {$t_m$} was varied in a broad range {$t_m\in\{1, 2,\dots,30\}t_c$}. The gradient strength is expressed in terms of the $b$-factor. For the filter block, it was {$b_f D_0\in \{1.13,\, 2.25,\, 3.38\}$} for all media; for the detection block, in experiments with isotropic crossing bars it was set to {$b D_0\in \{0,0.001, \dots, 0.1\}$}, while for anisotropic spaghetti, isotropic spaghetti, and packed spheres media it was {$b D_0\in \{0,0.01, \dots, 1\}$}. To simulate the FEXI measurement, we independently selected the gradient directions for mobility filter and the measurement from the {$x$}, {$y$}, and {$z$} directions for isotropic media, and from the eigenvector directions for anisotropic medium. Since the latter directions are time dependent, we used them for {$t=\Delta$} of the measurement block. Each simulation was conducted {$5$} times, giving the mean and the standard deviation shown in Figure \ref{fig_media_bar} and Figure \ref{fig_results}. The mean values were used to test the fitting ability of the commonly used exponential recovery of diffusivity \cite{lasivc2011apparent}, 
\begin{eqnarray}\label{def_fit}
\Dapp(t_m) = \left[1-a\,e^{-R\, t_m}\right] D_{\rm true}(\Delta) \,
\end{eqnarray}
and to roughly quantify the apparent exchange rate $R$. The value $D_{\rm true}(\Delta)$ is the true diffusion coefficient, Equation \ref{def_Dab} in the same direction as $\Dapp(t_m)$ and for the same diffusion time $\Delta$ as used in the FEXI measurement block (Figure \ref{FEXI_sequence}). Its value can be either taken from the MC result-- Equation \ref{def_Dab}, or fitted together with $a$ and $R$.

\section{Results}
All media are characterised with simulated time-dependent genuine diffusion tensor calculated according to Equation \ref{def_Dab}. For the isotropic crossing bars with the $1\times1$ voxel sides cross-section, the result agrees with the expected value $D_0/3$ (Figure \ref{fig_media_bar}), which is the fraction of spins in the bars parallel to a given direction, two other orientations contributing zero diffusivity. Increasing the bar cross-section results in the increase in the diffusion coefficient, also in the long-time limit. This is a consequence of diffusion-mediated exchange between the bars oriented parallel and perpendicular to the selected direction. 

Simulated FEXI results for isotropic crossing bars (Figure \ref{fig_media_bar_results}) show a pronounced dependence of diffusion coefficient on the mixing time. The simplicity of this media helps to illustrate the central idea: The spins inside the bars, which are orthogonal to the filter gradient, escape the dephasing by the mobility filter. During the mixing time, such spins leak to other bars (Figure \ref{fig_exchange}). When the diffusion measurement direction coincides with the filter direction, these ``fresh'' spins contribute the maximum diffusivity to the medium mean value, which results in the gradual recovery of diffusivity, Figure \ref{fig_media_bar_results}a. This is purely diffusion-mediated exchange, without any membrane permeation. 

The effect changes its sign, when the filter and measurement gradients are orthogonal,  Figure \ref{fig_media_bar_results}b. To explain that, the bars parallel to the measurement gradient contribute the maximum diffusivity right after the filter. As the mixing time increases, their weight in the overall signal decreases due to inflow of dephased spins, which results in an overall decrease in the diffusivity. For very long mixing time, the density of signal-contributing spins becomes uniform and the medium-averaged diffusivity recovers the unperturbed value. The signal is however contributed by nearly $2/3$ of all spins (in the absence of relaxation), which effectively increases the noise in the measurement results. 

The apparent exchange rate {$R$} was found by fitting Equation \ref{def_fit} to data. To support the idea presented in Figure \ref{fig_exchange}, we investigated the dependence of {$R$} on the segment length {$\ell$} of the medium constructed with thin bars, Figure \ref{fig_media_bar}. Since {$\ell$} is the only relevant parameters with the dimension of length, the characteristic time should scale as {$\ell^2/D_0$} and {$R$} and the inverse of it, {$R\propto 1/\ell^2$}. Figure \ref{fig_media_bar_results}c shows an excellent agreement with this scaling. Figure \ref{fig_media_bar_results}d shows a comparison of fitted $D_{\rm true}(\Delta)$ in Equation \ref{def_fit} with the really true value obtained from MC simulations using Equation \ref{def_Dab}. The good agreement between them reflects a good fit quality, which is also obvious from Figure \ref{fig_media_bar_results}a and b. The increase in diffusivity for small segment length reflects a more complex motion pattern than diffusion in straight channels. The characteristic segment length for which this increase is substantial can be estimated as $\ell_c = (D_0 \Delta)^{1/2}\approx 3\mbox{ voxels}$. 

Diffusion tensor eigenvalues for random media are shown in Figure \ref{fig_media}d--f. They confirm the strong anisotropy of the medium shown in panel (a), which we refer to as ``anisotropic sphaghetti''. The other spaghetti medium (panel (b)) is nearly isotropic, while the packed spheres are perfectly isotropic. Simulated FEXI results for these media are shown in Figure \ref{fig_results}. The most pronounced diffusion-mediated exchange is present for the most complex geometries (anisotropic spaghetti medium across the principal diffusion direction and the isotropic spaghetti medium). The magnitude of the effect and the exchange time decrease for simpler geometries (the principal direction in the anisotropic spaghetti and in packed spheres). Measurement in the direction orthogonal to the filter does not show a noticeable effect. 
An attempt to fit Equation \ref{def_fit} reveals a non-exponential approach of $D_{\rm app}(t_m)$ to its long-time value, Figure \ref{fig_results_fit}. While an approach to a common asymptote is seen in, e.g., Figure \ref{fig_results}a, the three-parameter fitting of Equation \ref{def_fit} does not catch this feature delivering different asymptotic values of $D_{\rm app}(t_m)$ for long $t_m$. 

\section{Discussion and Conclusions}
In this study, we investigated the imprint of diffusion-mediated exchange on FEXI. We have demonstrated the recovery of FEXI-derived apparent diffusion coefficient towards the unperturbed value in connected compartments of complex geometry where spins can find ``shelters'' in which the mobility of water molecules is restricted by geometrical constraints. Such spins escape the suppression by the mobility filter (Figure \ref{fig_exchange}) and contribute to the recovery of diffusion coefficient when moving in areas with less restricted diffusion. In contrast to previous simulation studies focused on packing of simple objects \cite{tian2017evaluation,ludwig2021apparent}, our main interest was on synthetic random media inspired by the complexity of brain gray matter, Figure \ref{fig_media}. The complex geometry of such media results in the emergence of the typical FEXI pattern of gradual recovery of diffusion coefficient towards its unperturbed value.

Obtained results challenge both the theory and the experiment. Theory of at least the simplest of considered media, the regular bars (Figure \ref{fig_media_bar}), would be useful to gain some intuition. The goal would be to find the long-time diffusion asymptote (Figure \ref{fig_media_bar}), the drop in the diffusivity right after the filter (Figure \ref{fig_media_bar_results}) and the apparent exchange time $R$. Note further the different functional forms of the diffusivity recovery for the regular bars and random media. The fast exponential recovery for the former is contrasted to the slow, presumably power-law recovery for the latter. This is alike the approach of genuine diffusion coefficient to its asymptote in regular and disordered media \cite{novikov2014revealing,Jespersen2019_ismrm}  although these results are not directly applicable to FEXI.  

On the experimental side, the main challenge is distinguishing the discussed diffusion-mediated effect and the genuine transcytolemmal exchange. Since the necessary theoretical basis is missing, we can only speculate that isotropic diffusion weighting \cite{mori1995single} might be useful removing the dependence on the direction of the filter gradient. It can be also helpful if the two effects had essentially different time scales, for example, when the diffusion-mediated exchange is fast and the membrane permeation is slow. Exploring the large parameter space of FEXI might help, in particular, studying the effect of mutual orientations of the filter and measurement gradients and the timing of diffusion-weighting gradients. 

Note that the presently used synthetic media are too loose models of biological cells, which suggests an obvious direction of future work.However, it is not easy to refrain from speculative projecting the present results on measurements in brain gray matter. Identifying the tube diameter with that of dendrites, $2\rho=1\,{\rm \mu m}$ and using $D_0=1\,{\rm \mu m^2/ms}$ gives for the time unit in Figure \ref{fig_results} $t_c=0.25\,{\rm ms}$. The exchange times about $10t_c$ turn then to very short values about $3\,{\rm ms}$. This figure is of course a very rough estimate. In particular, the exchange time $1/R$ is proportional to the squared segment length between bifurcations, Figure \ref{fig_media_bar_results}c, which means that longer segments would result in a higher values. Looking at available experimental data, we treat as the ground truth for the transcytolemmal exchange the results obtained in culture of vital neuronal cells giving the intracellular residence time of $750\pm 50\,{\rm ms}$ and $570\pm30\,{\rm ms}$ for neurons and astrocytes, respectively \cite{yang2018intracellular}. Similar values in the range $0.16-1.12\,{\rm s}$ were also obtained using \textit{in vivo} FEXI  \cite{nilsson2013noninvasive,bai2020feasibility,lasivc2016apparent}. Surprisingly, recent measurements exploiting other biophysical mechanisms provide evidences of much shorter exchange times when exchange is taken into account using anisotropic K\"arger model. So do the measurement at very strong diffusion weighting in which the signal is dominated by the intra-neurite contribution. Measurement up to $b=100\,{\rm ms/\mu m^2}$ require \textit{in vivo} exchange time of $10-30\,{\rm ms}$ to explain the deviation from the model of thin neurites \cite{veraart2020noninvasive}. Even faster exchange in brain gray matter is suggested when including in consideration the time dependence of diffusion-weighted signal. The obtained residence time is $2.5-3.6\,{\rm ms}$ \textit{in vivo} \cite{OLESEN2022118976} and $3.5\,{\rm ms}$ \textit{ex vivo} \cite{lee2022}. This agrees with the estimate of $10\,{\rm ms}$ obtained in excised neonatal mouse spinal cords using a $15\,{\rm T/m}$ static gradient in the stray field of a magnet \cite{williamson2019magnetic}. While the biophysical underpinning of these effects is still disputable, we cannot exclude a higher sensitivity of heavily diffusion weighted signal to the fast diffusion-mediated exchange rather than the slow transcytolemmal permeation. 

The discussed geometrical effects may be present in intensively branched dendritic tree, as in Purkinje or granule cells, or due to exchange between soma and processes, as in astrocytes \cite{bressloff1996asum,grebenkov2005diffusion}. In general, diffusion on ramified trees challenges researchers since decades. Diffusion-weighted NMR has been also considered focusing on the time-dependent diffusion \cite{van2015numerical,palombo2016new} and in the context of double diffusion encoding \cite{ianus2021mapping}, which is a close approach to FEXI. 

Summarizing this study with an answer to the question in the title, FEXI does measure exchange, but exchange mediated by both the membrane permeation and diffusion within individual compartments of tortuous geometry in agreement with previous studies.  The role of the latter in biomedical measurements remains to be clarified.

\section*{Acknowledgments}
We are grateful to Dmitry Novikov, Marco Palombo, Denis Grebenkov, Saad Jbabdi, and numerous members of microClub ($\mu$Club) for constructive discussions of this study. This work  was supported by Academy of Finland (grant \#323385 to A.S.) and Erkko Foundation (A.S.). We would like to thank Bioinformatics Center at University of Eastern Finland, Finland, for computational resources. 

\bibliography{wileyNJD-AMA} 

\end{document}